\documentclass[prl,showpacs,twocolumn,superscriptaddress]{revtex4-1}

\usepackage{graphicx}% Include figure files
\usepackage{bm}% bold math
\usepackage{amsmath}
\usepackage{amsfonts}

\newcommand{\vep}{\varepsilon}

\begin{document}

% Title and abstract

\title{Semiconductor-inspired design principles for superconducting quantum computing}

 \author{Yun-Pil Shim}
 \email{ypshim@lps.umd.edu}
 \affiliation{Laboratory for Physical Sciences, College Park, Maryland 20740, USA}
 \affiliation{Department of Physics, University of Maryland, College Park, Maryland 20742, USA}
 \author{Charles Tahan}
 \email{charlie@tahan.com}
 \affiliation{Laboratory for Physical Sciences, College Park, Maryland 20740, USA}
 \date{\today}

% Abstract
\begin{abstract}
Superconducting circuits offer tremendous design flexibility in the quantum regime culminating most recently in the demonstration of few qubit systems supposedly approaching the threshold for fault-tolerant quantum information processing. Competition in the solid-state comes from semiconductor qubits, where nature has bestowed some very useful properties which can be utilized for spin qubit based quantum computing. Here we begin to explore how selective design principles deduced from spin-based systems could be used to advance superconducting qubit science. We take an initial step along this path proposing an encoded qubit approach realizable with state-of-the-art tunable Josephson junction qubits. Our results show that this design philosophy holds promise, enables microwave-free control, and offers a pathway to future qubit designs with new capabilities such as with higher fidelity or, perhaps, operation at higher temperature. The approach is also especially suited to qubits based on variable super-semi junctions.
\end{abstract}

\maketitle

% Main body

%\section{Introduction}

Spin qubits \cite{spin_qubit_review} are based on the fundamental and intrinsic properties of semiconductor systems, such as electron spins trapped in the potential of a quantum dot \cite{loss_divincenzo_pra1998} or a chemical impurity \cite{kane_nature1998}. Spins can be naturally protected from charge noise due to weak spin-orbit coupling. In fact, the tiny matrix element between spin qubit states can allow spin qubits to operate at temperatures above the Zeeman splitting \cite{Tyryshkin_Lyon_nmat2012,Saeedi_Thewalt_science2013}. While a benefit to qubit coherence, this property of spins also leads to relatively slow single qubit gates via, for example, a microwave pulse. It turns out that nature provides a solution: a very fast and robust two-qubit gate via the exchange interaction. This has led to ``encoded'' qubit schemes where the qubit is embedded logically in two to four physical spin qubits \cite{Petta2005,divincenzo_bacon_nature2000,bacon_whaley_prl2000}. The fact that electrons are real particles can be used for fast initialization and readout techniques. Exchange-only qubits \cite{divincenzo_bacon_nature2000,fong_wandzura_qic2011} allow all electrical implementation of qubit gate operations and enable universal quantum computation (QC) while providing some immunity to global field and timing fluctuations via a decoherence free subsystem, at the cost of more physical qubits and extra operations per encoded gate. 

This work investigates how superconducting (SC) Josephson junction quantum circuits \cite{SC_qubit_review}, whose properties can be engineered, can be improved by mimicking some of best properties of spin qubit systems. We propose a first step: an encoded superconducting qubit approach which does not require microwave control, and thus divorces qubit frequency from control electronics. In analogy to the exchange only qubit in semiconductor spin qubit systems, encoded qubits enable microwave-free control of the qubit states via fast DC-like voltage or flux pulses. In contrast to the exchange only qubits, logical gate operations of this encoded SC qubit can be done with minimal overhead (zero overhead in physical 2-qubit gates) in terms of control operations, a surprising result. We describe how to initialize the encoded qubit and implement single- and two-qubit logical gates using only $z$-control pulse sequences (via tunable frequency qubits). In the process we also lay out possible opportunities for future research based on other insights from spin-based QC. To encourage implementation we give an explicit protocol based on qubits in operation today.

Small systems of superconducting qubits based on variations of the transmon qubit \cite{transmon,transmon_exp} have already demonstrated gates with fidelities approaching 99.99\% along with rudimentary quantum algorithms including error correction cycles \cite{xmon,5xmon,9xmon,Reed_Schoelkopf_nature2012, Sun_Schoelkopf_nature2014,Chow_Steffen_ncomm2014,Corcoles_Chow_ncomm2015}. Note that because these architectures rely on single qubit gates via microwaves, the future design space is constrained by the availability and convenience of microwave generators. 

An alternative approach to combining the best properties of semiconductor and superconducting quantum systems is to take advantage of true superconducting-semiconductor systems. The appearance of superconductivity in conventional semiconductors \cite{blase_nmat2009,Bustarret_PhysicaC2015} such as silicon \cite{bustarret_nature2006,marcenat_prb2010,dahlem_prb2010,grockowiak_sst2013} or germanium \cite{herrmannsdorfer_prl2009,skrotzki_ltp2011} could potentially allow for a new type of fully epitaxial super-semi devices \cite{shim_tahan_ncomm2014,shim_tahan_ieee2015}. And epitaxial super-semi Josephson juction devices based on the proximity effect have already led to new superconducting circuits \cite{nanowireJJ_Marcus,nanowireJJ_DiCarlo}. Epitaxial super-semi systems may improve noise properties, but perhaps more importantly they enable gate-tunable Josephson junctions, which we can also take advantage of in our proposal introduced below.

\section{Results}

\subsection{From encoded spins to tunable qubits}

% Fig1
\begin{figure}
  \includegraphics[width=\linewidth]{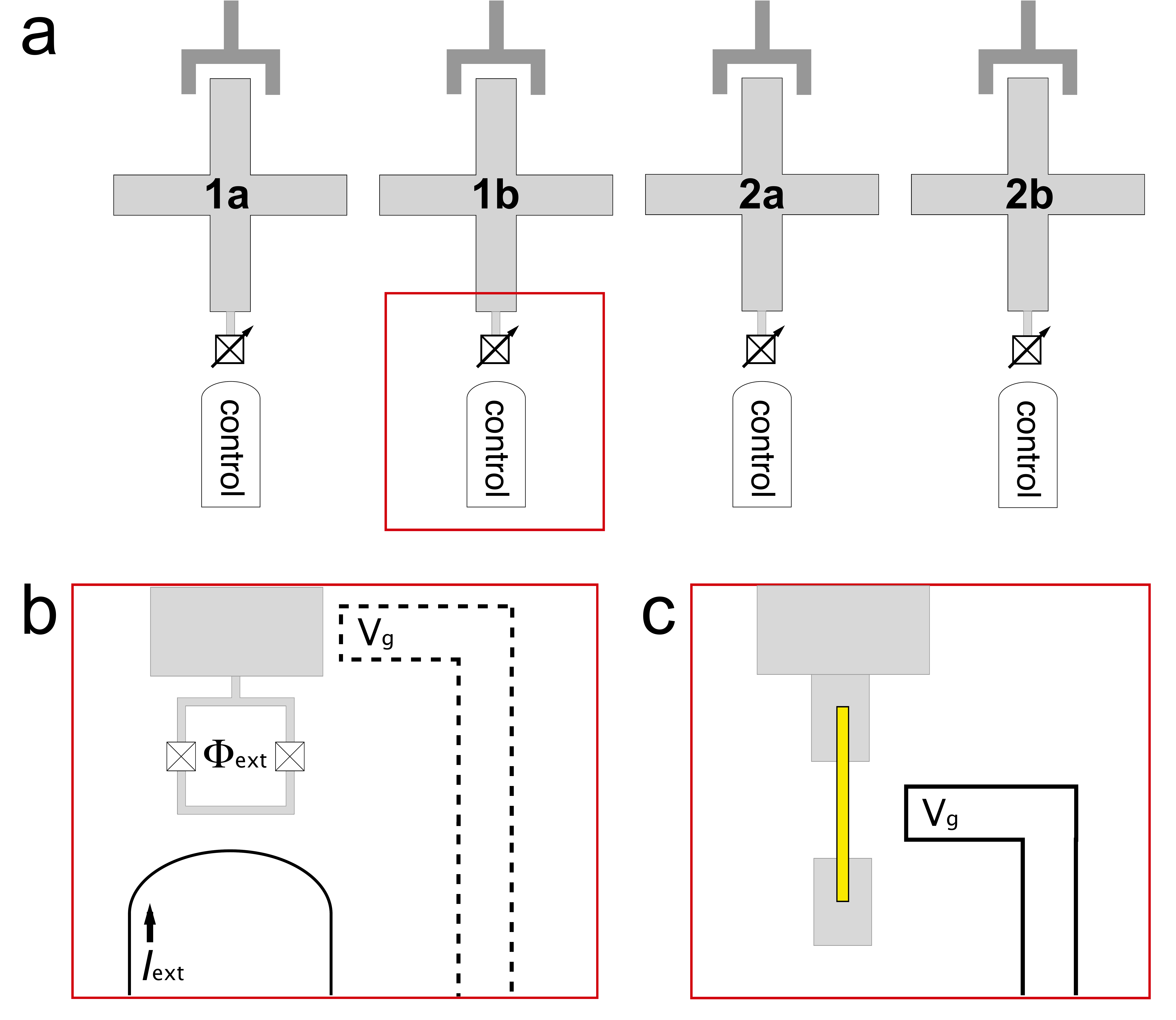}\\
  \caption{{\bf Encoded superconducting qubits and tunable Josephson junctions.} {\bf (a)}, Schematic diagram of a possible encoded superconducting qubit scheme as described in the text. An encoded qubit consists of two tunable physical SC qubits ({\it e.g.} tunable transmons such as the xmon depicted here), with the encoded qubit states $|0\rangle_{\mathrm Q}$=$|01\rangle$ and $|1\rangle_{\mathrm Q}$=$|10\rangle$. In this picture, two encoded qubits are shown (e.g., physical qubits 1a and 1b form an encoded qubit) and more encoded qubits can be introduced in a straightforward manner. Each SC qubit has a $z$-control line which tunes the Josephson energy $E_{\mathrm J}$, and there are no additional microwave $xy$-control lines. All manipulation of the qubit states are done by the $z$-control pulses. Each transmon is capacitively coupled to neighboring transmons and also coupled to, as an example, a transmission line resonator for readout. {\bf (b)}, double JJs in a loop act as a tunable JJ, controlled by an externally applied magnetic flux. In the SQUID tunable approach, one of the transmons in each encoded qubit needs a separate voltage control to tune the gate charge number $n_{\mathrm g}$ which may be needed to initialize the encoded qubit state. {\bf (c)}, electrostatically tunable JJ based on a proximitized superconducting-semiconductor nanowire connecting two superconductors \cite{nanowireJJ_Marcus} used for gatemons. The nanowire is coated with SC and a portion is lifted off to form a semiconductor nanowire weak link. The JJ energy $E_{\mathrm J}$ is tuned by a side-gate voltage $V_{\mathrm G}$, which can also serve as capacitive tuning for initialization.} 
  \label{fig:system}
\end{figure}

Spins in quantum dots, say in silicon, are typically assumed to have equivalent g-factors, so that in a magnetic field the frequency of each qubit is the same. Thus, to achieve universal quantum computation (an ability to do arbitrary rotations around the Bloch sphere plus a two-qubit entangling gate), one needs at minimum three spins. In this case, an encoded 1-qubit gate requires around 3 pulses and a CNOT gate requires roughly 20 pulses \cite{divincenzo_bacon_nature2000}, a hefty overhead. 2-qubit encodings are possible, but require the complication of a magnetic field gradient (via for example a micromagnet). Superconducting qubits, on the other hand, can be man-made such that the qubit frequency is tunable. This allows arbitrary one-qubit rotations with just two physical qubits, in theory. 

In this work, we consider a qubit encoded in a system of two capacitively coupled SC qubits.
We take tunable transmons \cite{Reed_Schoelkopf_nature2012,Riste_DiCarlo_ncomm2015} like xmons \cite{xmon} or gatemons \cite{nanowireJJ_Marcus} as our prototypical SC qubits (see Fig. \ref{fig:system}a) and suggest one possible implementation following the capacitively-coupled xmon architecture of Martinis et al. \cite{5xmon} to encourage near-term realization.
Although we explicitly chose the xmon geometry to be more specific about our proposed protocol, the general idea can easily be applied to other types of SC qubits, such as traditional transmons or capacitively-shunted flux qubits \cite{fluxmon_theory,fluxmon_Oliver}, which we will discuss later. 
A transmon qubit \cite{transmon} is described by the charge qubit Hamiltonian
\begin{equation}
H_\mathrm{X} = 4 E_{\mathrm C} \left( \hat{n} - n_{\mathrm g} \right)^2 - E_{\mathrm J} \cos\hat{\theta} ~, 
\end{equation}
where $E_{\mathrm C}$=$e^2/2C_{\Sigma}$ is the electron charging energy for total capacitance $C_{\Sigma}$ and $E_{\mathrm J}$ is the Josephson energy. 
$\hat{n}$ and $\hat{\theta}$ are the number and phase operators, respectively, and $n_{\mathrm g}$ is the gate charge number that can be tuned by a capacitively-coupled  external voltage. The qubit frequency $f_{\mathrm Q}$=$\vep/h$ where $\vep$ is the energy difference between the first excited state and the ground state, and $f_{\mathrm Q} \simeq \sqrt{8 E_{\mathrm C} E_{\mathrm J}}/h$ in the transmon regime, $E_{\mathrm J} \gg E_{\mathrm C}$. 
The Josephson energy of a JJ is determined by the material properties and geometry of the JJ, but a double JJ can be considered as a tunable JJ \cite{Makhlin_tunableJJ} where an externally applied magnetic flux through the double JJ loop can tune the effective coupling energy $E_{\mathrm J}$=$E_{J0} \cos\left( \pi \Phi_\mathrm{ext}/\Phi_0 \right)$ (see Fig. \ref{fig:system}b). $\Phi_\mathrm{ext}$ is the external magnetic flux and $\Phi_0$ is the SC flux quantum.   
Individual transmon qubits are typically controlled by tuning the qubit frequency with tunable $E_{\mathrm J}$ for $z$ control and by applying microwaves for $x$ control. 

Recently, there has been progress in an alternative approach for a tunable JJ using a superconductor proximitized semiconductor weak link junction \cite{nanowireJJ_Marcus,nanowireJJ_DiCarlo}. In Ref. \cite{nanowireJJ_Marcus}, an InAs nanowire was used to connect two superconductors (Al). The nanowire was epitaxially coated with Al and a small portion of the wire was etched off to form a semiconductor nanowire bridging two SCs (Fig. \ref{fig:system}c). A side-gate voltage was used to tune the carrier density under the exposed portion of the wire and thus the Josephson energy of this SNS JJ. The gatemon, a tunable transmon based on this gate-tunable JJ, has several advantages. It requires only a single JJ that can be quickly tuned by a electrostatic voltage. It removes the need for external flux, and hence reduces dissipation by a resistive control line and allows the device to operate in a magnetic field. The epitaxial growth of the nanowire JJ and its clean material properties \cite{Chang_Marcus_nnano2015,Krogstrup_Jesperson_nmat2015} demonstrate the potential of a bottom-up approach for SC quantum devices \cite{shim_tahan_ncomm2014,shim_tahan_ieee2015}. 

Our encoded qubit is defined in a two-transmon system.
The Hamiltonian for two transmons with the capacitive $xx$ coupling is 
\begin{eqnarray}\label{eq:H2X}
H_\mathrm{2X} &=& \sum_{k=a,b} \left[ 4 E_{\mathrm C}^{(k)} \left( \hat{N}_k - n_{\mathrm g}^{(k)} \right)^2 - E_{\mathrm J}^{(k)} \cos\hat{\theta}_k \right] \nonumber\\
              &&  + E_{\mathrm{cc}} \left( \hat{N}_a - n_{\mathrm g}^{(a)} \right) \left( \hat{N}_b - n_{\mathrm g}^{(b)} \right) \nonumber\\
&=& \vep_a \tilde{\sigma}_a^z + \vep_b \tilde{\sigma}_b^z + \vep' \tilde{\sigma}_a^x  \otimes \tilde{\sigma}_b^x ~,
\end{eqnarray} 
where $E_{\mathrm{cc}}$ is the capacitive coupling energy and $\tilde{\sigma}_k^i$ ($i$=$x,y,z$) is the Pauli operator for the $k$-th transmon in a reduced subspace of transmon qubit states. $\vep_k$ is the qubit energy of the $k$-th transmon, and $\vep'$=$E_{\mathrm{cc}} \alpha_a \alpha_b$ with $\alpha_k$=$\langle 1 | \hat{N}_k | 0\rangle$ 
where $|0\rangle$ and $|1\rangle$ are the two lowest energy states of individual transmons.  
In transmon qubit systems, the capacitive coupling is usually turned on (off) by tuning the qubit frequencies to on (off) resonance.
The capacitive $xx$ coupling conserves the parity $\tilde{\sigma}_a^z  \otimes \tilde{\sigma}_b^z$ of the two transmon system, 
and the Hamiltonian [Eq. (\ref{eq:H2X})] is block-diagonal in the basis of 
$\{ |00\rangle, |01\rangle, |10\rangle, |11\rangle \}$.
We define our encoded qubit in the subspace of $\langle \{ |01\rangle,  |10\rangle \} \rangle$, 
since the other subspace $\langle \{ |00\rangle,  |11\rangle \} \rangle$ has states with a very large energy difference (much larger than the capacitive coupling), effectively turning off the capacitive coupling all the time.

In the encoded qubit basis $\{ |0\rangle_{\mathrm Q}, |1\rangle_{\mathrm Q} \} $ where 
\begin{equation}\label{eq:logicalQ}
|0\rangle_{\mathrm Q} = |01\rangle, \quad\quad |1\rangle_{\mathrm Q} = |10\rangle ~,
\end{equation}
the single qubit Hamiltonian is
\begin{equation}\label{eq:HQ}
H_\mathrm{Q} = \left( 
\begin{array}{cc}
- \vep_a + \vep_b & \vep' \\
\vep' & \vep_a - \vep_b 
\end{array}
\right)
=\frac{\vep_a+\vep_b}{2} \openone + \Delta\vep \hat{\sigma}^z + \vep' \hat{\sigma}^x ~,
\end{equation}
where $\Delta\vep$=$(\vep_b-\vep_a)/2$ and $\hat{\sigma}^i$ ($i$=$x,y,z$) is the Pauli operator for the encoded qubit.
The qubit energies $\vep_a$ and $\vep_b$ can be controlled by the tunable JJ of each tunable transmon or gatemon, enabling logical gate operations with only fast DC-like voltage or flux pulses. In the following we will describe the logical gate operations, initialization, and measurement schemes for this encoded qubit architecture.

\subsection{Single qubit operations}

% Fig2
\begin{figure}
  \includegraphics[width=\linewidth]{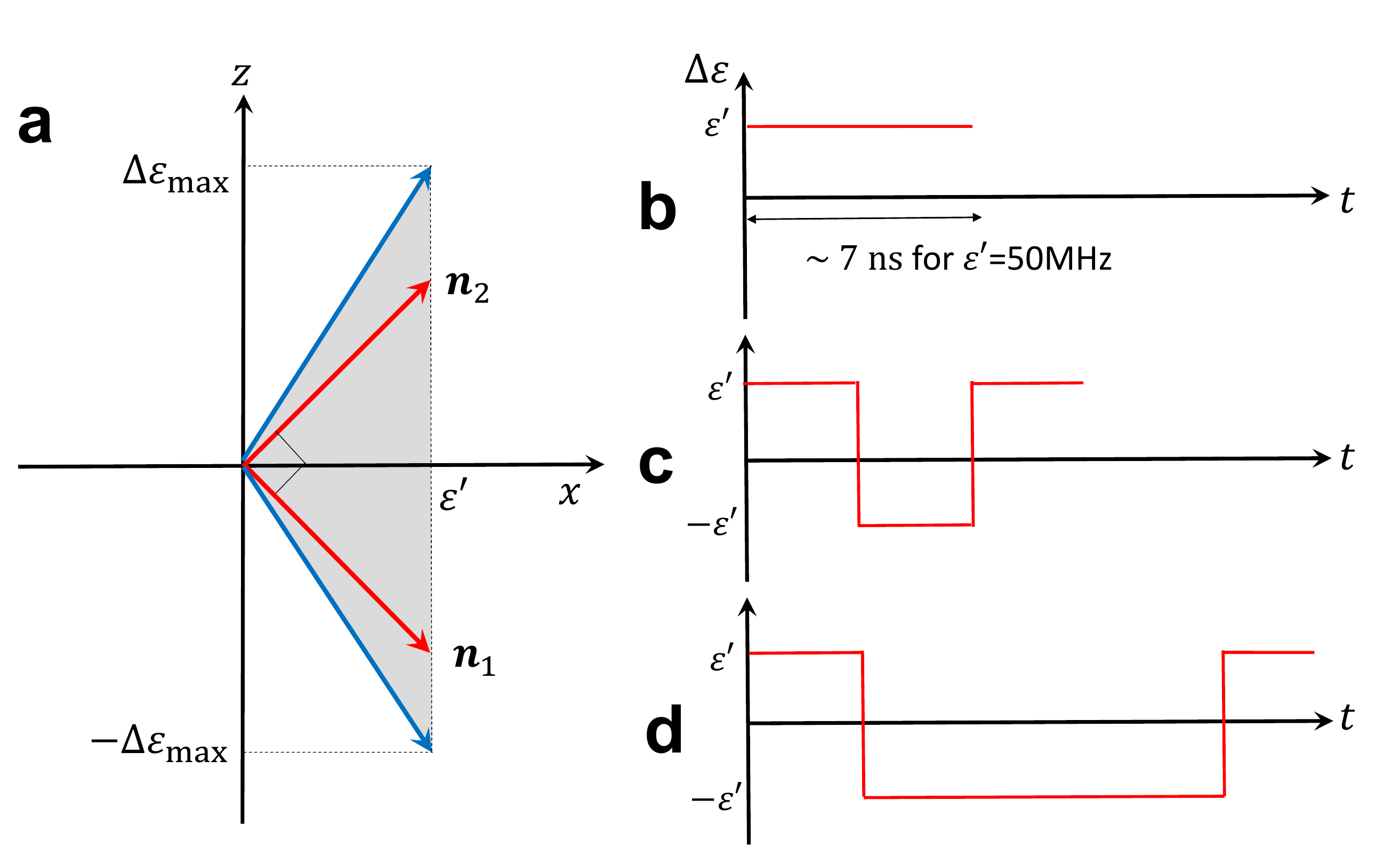}\\
  \caption{{\bf Single qubit operations.} {\bf (a)}, Possible rotation axes in $xz$ plane. The shaded gray region depicts the range of the direction of the possible rotation axis. The two red directions indicate a set of two orthogonal rotation axes, which can be used to implement any arbitrary single qubit gates in three steps. {\bf (b),(c),(d)} schematically shows an implementation of some logical gates in terms of rotations around $\hat{\mathbf n}_1$=$\mathbf{n}_1/|\mathbf{n}_1|$ and $\hat{\mathbf n}_2$=$\mathbf{n}_2/|\mathbf{n}_2|$. {\bf (b)}, Pulse shape for Hadamard gate. {\bf (c)}, Pulse shape for $X$ gate. {\bf (d)}, Pulse shape for $Z$ gate.}
  \label{fig:1Q_gates}
\end{figure}

The Hamiltonian for an encoded qubit is given by Eq. (\ref{eq:HQ}).
For a fixed capacitive coupling between SC qubits, $\vep'$ is fixed, and the single qubit operations can be implemented by pulsing the qubit energy $\vep$ through the $z$-control of individual transmons, in at most three rotations. 
Since the tunable range of $\Delta\vep$ (order of GHz) is much greater than $\vep'$ (tens or hundreds of MHz), the rotation axis can be in almost any direction in the right half of the $xz$ plane (see Fig. \ref{fig:1Q_gates}a), and most logical single qubit gates can be implemented in two rotations \cite{two_step_sq_gate}. 
In general, all single qubit gate operations can be implemented as a three-step Euler angle rotations around two orthogonal rotation axes ({\it e.g.} see the two red axes in Fig. \ref{fig:1Q_gates}a).

We now provide implementations for a few representative single qubit gates.
The Hadamard gate, $H$=$((1,1),(1,-1))/\sqrt{2}$, is a single qubit gate that is almost ubiquitous in quantum circuits. 
Figure \ref{fig:1Q_gates}b shows an implementation of $H$ gate as a single rotation $H$=$i R(\hat{\mathbf n}_2,\pi)$ around $\hat{\mathbf n}_2$=$(1,0,1)/\sqrt{2}$.
It can be achieved by tuning $\delta\vep$=$\vep'$. 
Here $R(\hat{\mathbf n},\phi)$ is a rotation by angle $\phi$ around $\hat{\mathbf n}$ axis.
Pauli $X$ gate can be realized as a single rotation by tuning the two xmons on resonance ($\Delta \vep$=0), or three-step rotations such as
$X$=$i R(\hat{\mathbf n}_2,\pi/2) R(\hat{\mathbf n}_1,\pi/2) R(\hat{\mathbf n}_2,\pi/2)$ where $\hat{\mathbf n}_1$=$(1,0,-1)/\sqrt{2}$ and $\hat{\mathbf n}_2$=$(1,0,1)/\sqrt{2}$, 
as was shown in Fig. \ref{fig:1Q_gates}c.
$Z$ gate requires three-step rotations: $Z$=$-i R(\hat{\mathbf n}_2,\pi/2) R(\hat{\mathbf n}_1,3\pi/2) R(\hat{\mathbf n}_2,\pi/2)$.
The above examples are for ideal systems with precise control over the system parameters. In real systems with fluctuating parameters, recently developed dynamical error-cancelling pulse sequences \cite{wang_bishop_ncomm2012,wang_bishop_pra2014} could be useful for gate operations with higher fidelity. 

Given that single qubit gates in transmon systems through $z$-control have already demonstrated fidelities better than 0.999 \cite{5xmon}, we expect the logical single qubit gates (which require at most three rotation steps through $z$ control of transmons) will be able to reach a fidelity better than $F_1 \ge F_z^3 = 0.999^3$=0.997 using currently available experimental techniques.

\subsection{Two qubit operations}

For a scalable qubit architecture, we need to plan for the transmon qubit frequencies such that unnecessary resonances are avoided, especially if the two-qubit interaction cannot be completely shut off via, e.g., a tunable coupler \cite{gmon}.
An encoded qubit has two transmons with idle frequency difference much larger than the capacitive coupling, so we can effectively turn the coupling off. 
In the two encoded qubit system (4 transmon system), we set the idle frequencies of next-nearest neighbor transmons to be different by more than the direct capacitive coupling between them, which is order of MHz \citep{9xmon}. We also set the encoded qubit frequencies $\Delta\vep$ of the neighboring encoded qubits to be different so we can mitigate some unintended resonances. For the calculations in this section, we set the four transmon qubit idle frequencies $f_{\mathrm Q}^{(k)}$ as 5.6, 4.6, 5.9, 4.8 GHz for $k$ = 1a, 1b, 2a, 2b, respectively [see Fig. \ref{fig:system}a]. 
In this section and the following, we set $E^{(k)}_{\mathrm C}/h$=375MHz and $E_{\mathrm{cc}}$=30MHz for all transmons. Transmon qubit frequencies are controlled by tuning $E^{(k)}_{\mathrm J}$.

% Fig3
\begin{figure}
  \includegraphics[width=\linewidth]{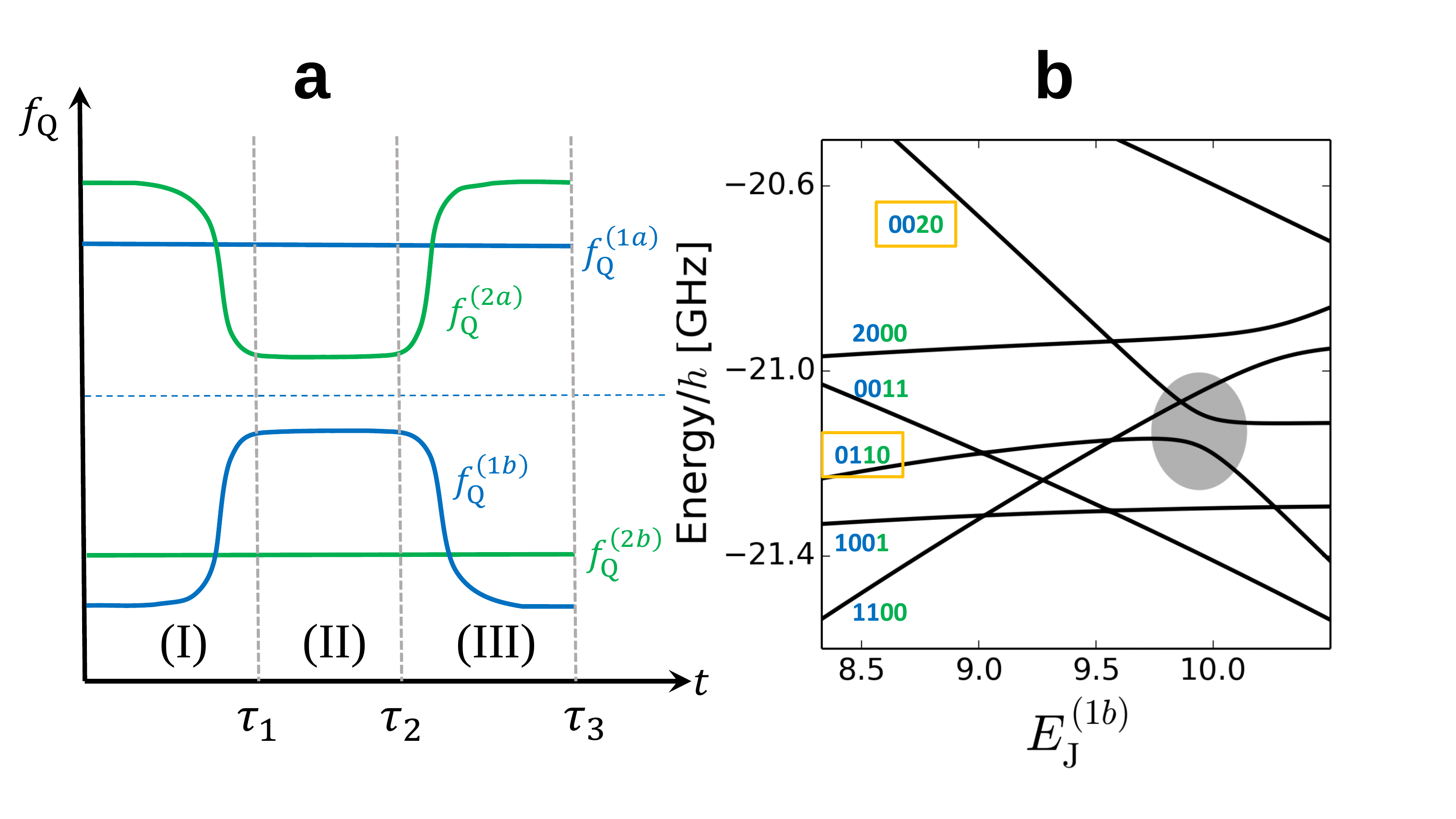}\\
  \caption{{\bf Two-qubit gate operation.} {\bf (a)}, Pulse scheme for two qubit gate operations. $y$ axis ($f_{\mathrm Q}$) is the qubit frequencies of each transmons in Fig. \ref{fig:system}a. $f_{k}^{(0)}$ is the idle qubit frequency of $k$-th transmon. The two blue curves (transmons 1a and 1b) form an encoded qubit, and the two green curves (transmons 2a and 2b) form the other encoded qubit. The transmons 1b and 2a are brought close to resonance while still far from being resonant with other transmons (transmon 1a and 2b), then are brought back to respective idle frequencies. {\bf (b)}, Energy spectrum for the process. The system is brought to the shaded area where (0110) and (0020) states are mixed. (0110) state accumulates nontrivial phase during this process, which leads to a CPHASE gate between transmon 1b and transmon 2a. This provides a non-trivial two-qubit gate necessary for universal QC.}
  \label{fig:two_qubit_cphase}
\end{figure}

Two qubit operations can be implemented by adopting the adiabatic two-qubit CPHASE operations \cite{dicarlo_chow_nature2009,5xmon} between two transmon qubits. 
By tuning the qubit frequencies of two transmon qubits such that (11) and (02) states become resonant and then bringing them back to their idle frequencies, a unitary gate equivalent to the CPHASE gate between two qubits up to single qubit unitary gates can be achieved \cite{adiabatic_cphase_Wellstood_PRL2003}. This scheme has already been used in experiments and achieved reported fidelity better than 0.99 \cite{5xmon}. 
In a similar manner, we can implement the CPHASE gate between two encoded qubits up to single qubit unitary gates. 
Figure \ref{fig:two_qubit_cphase}a shows schematically the pulse sequence of the transmon qubit frequencies, 
changing the qubit frequencies of transmon 1b and transmon 2a in Fig. \ref{fig:system}a. 
First, we bring the transmons 1b and 2a closer during time $\tau_1$ such that (0110) and (0020) states are on resonance in step (I). Then, in step (II),  they stay there for a time period $\tau_{12}$=$\tau_2-\tau_1$, and finally we bring them back to initial point at time $\tau_3$=$\tau_2+\tau_1$ in step (III). 
The (0110) state gets mixed with (0020) due to the capacitive coupling during the pulse sequence with strength $\sqrt{2} \vep'$.
During this process the (0110) state obtains some nontrivial phase due to the interaction with (0020) while the other qubit states, (0101), (1001), and (1010), obtain only trivial phases since they don't get close to any other states that can mix. 
This process results in a unitary operation in the encoded qubit space, up to a global phase, 
\begin{equation}
U = \left( 
\begin{array}{cccc}
e^{i\phi_2} & 0 & 0 & 0 \\
0 & e^{i \left( \phi_2 + \phi_3 + \delta\phi \right)} & 0 & 0 \\
0 & 0 & 1 & 0 \\
0 & 0 & 0 & e^{i\phi_3}  
\end{array}
\right) ~.
\end{equation}
This is equivalent to the CPHASE gate $(1,1,1,e^{i\delta\phi})^T$ up to single qubit operations.
\begin{eqnarray}\label{eq:UCPHASE}
{\mathrm{CPHASE}} &=& \left[ \left( \begin{array}{cc} 0 & 1 \\ e^{-i\phi_2} & 0 \end{array} \right) 
            \otimes \left( \begin{array}{cc} 1 & 0 \\ 0 & 1 \end{array} \right) \right] \nonumber\\
&& \times U
\times \left[ \left( \begin{array}{cc} 0 & 1 \\ 1 & 0 \end{array} \right) 
       \otimes \left( \begin{array}{cc} 1 & 0 \\ 0 & e^{-i\phi_3}  \end{array} \right) \right] ~.
\end{eqnarray} 

Note that, unlike Ref. \cite{dicarlo_chow_nature2009,5xmon}, we tune both transmons 1b and 2a instead of tuning only one of them. If we only tuned transmon 2a to bring the (0110) state close to the (0020) state, then transmon 2a and transmon 2b would be close to resonance. Because the transmon-transmon interaction through capacitive coupling can be turned on and off by bringing the transmons on and off resonance, this will result in a complicated, unintended operation as well as leakage. 
So it is necessary to tune transmons 1b and 2a simultaneously so that transmons 1a and 2b do not come into play during the process. The resonance between next-nearest neighbors can also lead to some small anti-crossing, but these resonances only occur during the fast ramping up and down steps and thus can be negligible.
This scheme is preferable to directly using the $xx$ coupling between transmons 1b and 2a, since $xx$ coupling drives the system outside of the encoded qubit space and hence leads to leakage, requiring a rather long sequence of pulse gates to implement a two-qubit logical operation \cite{divincenzo_bacon_nature2000,fong_wandzura_qic2011}. 
The physical CPHASE gate has been successfully implemented for xmon qubits with gate time of $\sim$ 40 ns \cite{5xmon}, which can be directly applied for logical two-qubit gate here, too.

% Fig4
\begin{figure}
  \includegraphics[width=\linewidth]{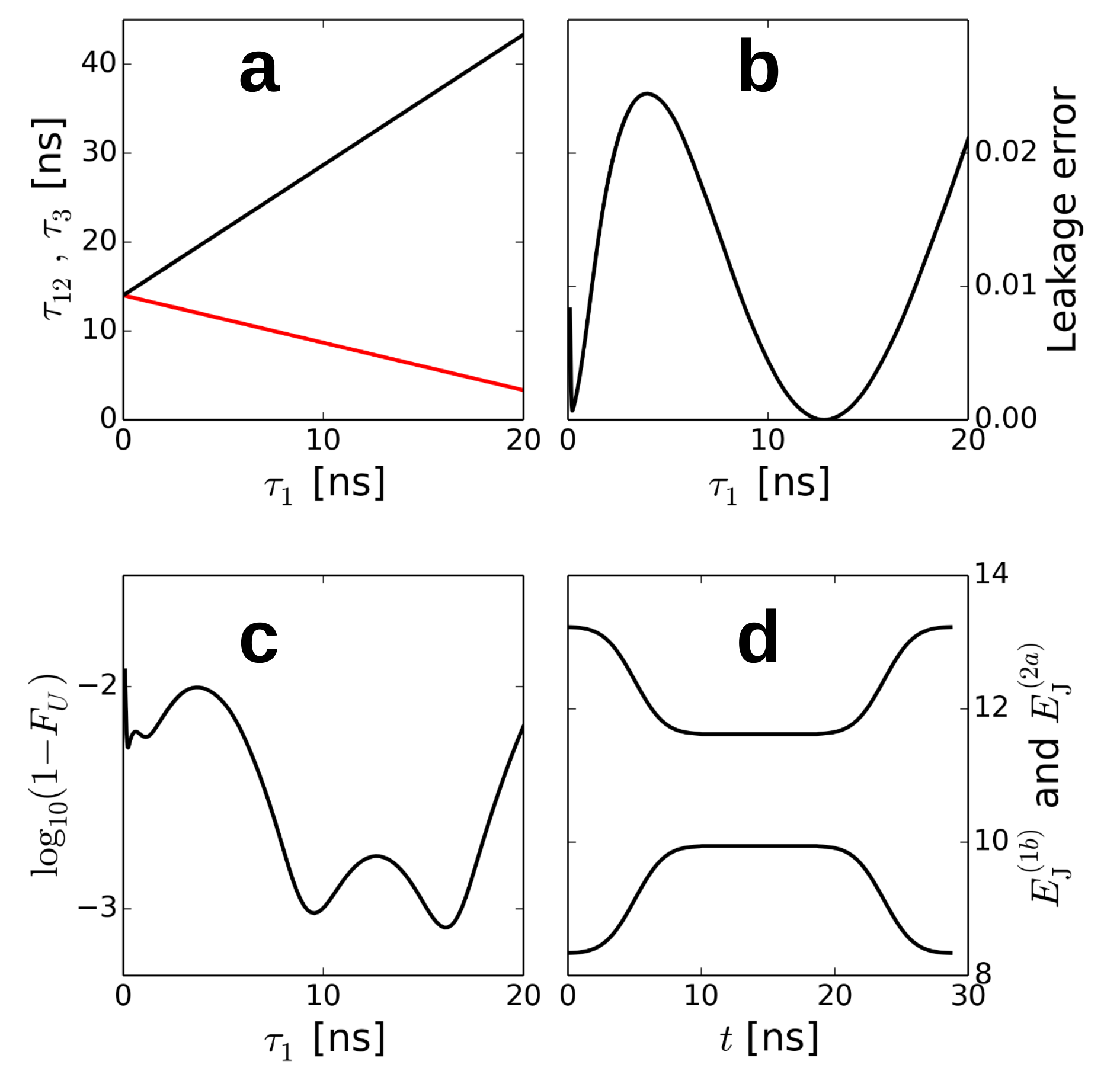}\\
  \caption{{\bf Fidelity of adiabatic CZ interaction operation.} {\bf (a)}, Operation time. The red curve is the staying time $\tau_{12}$ = $\tau_2-\tau_1$ and the black curve is the total time $\tau_3$.  {\bf (b)}, Leakage error during the process. {\bf (c)}, Gate fidelity in terms of Makhlin invariants. This gives a measure of how close the unitary gate is to the CZ gate up to single qubit operations. {\bf (d)}, Pulse shape for $\tau_1$=10ns. We used error function to model a smooth pulse shapes for $E_{\mathrm J}^{(1b)}$ and $E_{\mathrm J}^{(2a)}$.}
  \label{fig:cphase_fidelity}
\end{figure}

Figure \ref{fig:cphase_fidelity} shows simulated numerical results of this physical two qubit interaction between transmons 1b and 2a. We use an error function shape ramping up and down, similar to Ref. \cite{ghosh_geller_pra2013}, 
\begin{eqnarray}
&& E_{\mathrm J}^{(1b)} (t)  \nonumber\\
&& = \left\{ 
\begin{array}{ll}
E_{\mathrm J0}^{(1b)} + \frac{E_{\mathrm{res}}^{(1b)} - E_{\mathrm J0}^{(1b)}}{2} \left( 1 + \mathrm{erf}\left( \frac{t - \tau_1/2}{ \sqrt{2} \sigma} \right) \right) & (\mathrm{I}) \\
E_{\mathrm{res}}^{(1b)}   & (\mathrm{II}) \\
E_{\mathrm{res}}^{(1b)} - \frac{E_{\mathrm{res}}^{(1b)} 
    - E_{\mathrm J0}^{(1b)}}{2} \left( 1 + \mathrm{erf}\left( \frac{t - \tau_1/2 - \tau_2}{ \sqrt{2} \sigma} \right) \right)  & (\mathrm{III}) 
\end{array}
\right. %\nonumber
\end{eqnarray}
and $E_{\mathrm J}^{(2a)} (t)$= $E_{\mathrm J0}^{(1b)} + E_{\mathrm J0}^{(2a)} - E_{\mathrm J}^{(1b)} (t)$. 
$E_{\mathrm J0}$ is the idle value and $E_{\mathrm{res}}$ is for resonant (11) and (02) states.
To find optimal solutions of this form, we change $\tau_1$ and choose $\sigma$=$\tau_1/4\sqrt{2}$. $\tau_{12}$=$\tau_2-\tau_1$ is calculated analytically using a perturbative expression such that the whole process will result in the $U$ with desired $\delta\phi$.
Figure \ref{fig:cphase_fidelity}a shows $\tau_{12}$ and the total time $\tau_3$ needed to implement a CZ gate ($\delta\phi=\pi$). 

Due to the mixing with higher energy states which are out of the encoded qubit space, leakage error could pose a problem.
We can compute the leakage error as follows. 
The full unitary operation matrix $U$ can be written in a block-form
\begin{equation}
U = \left( \begin{array}{cc} U_\mathrm{AA} & U_\mathrm{AB} \\ U_\mathrm{BA} & U_\mathrm{BB} \end{array} \right)~,
\end{equation}
where $\mathrm{A}$ is the encoded qubit subspace and $\mathrm{B}$ is the complementary subspace.
For any qubit state $|\psi_\mathrm{A}\rangle$ in the encoded qubit space, the leaked portion is $U_\mathrm{BA} |\psi_\mathrm{A}\rangle$ and 
$||U_\mathrm{BA} |\psi_\mathrm{A}\rangle||^2$=$\langle \psi_\mathrm{A} | U^{\dag}_\mathrm{BA} U_\mathrm{BA} |\psi_\mathrm{A}\rangle \equiv \langle \psi_\mathrm{A} | W_\mathrm{AA} |\psi_\mathrm{A}\rangle$.
$W_\mathrm{AA}$=$U^{\dag}_\mathrm{BA} U_\mathrm{BA}$ is positive definite and the leakage error $E_{\mathrm{leak}}$ can be defined as $\mathrm{max} \langle \psi_\mathrm{A} | W_\mathrm{AA} |\psi_\mathrm{A}\rangle$ = $\mathrm{max}_{\lambda}\{w_{\lambda}\}$ where $w_{\lambda}$ are the eigenvalues of $W_\mathrm{AA}$.    
The leakage error (Fig. \ref{fig:cphase_fidelity}b) can be a few percent, but if we choose optimal $\tau_1$, it can be significantly reduced, well below 1\%. 
Note too that leakage can be dealt with algorithmically \cite{Wu_Lidar_leakage_prl2002,Fowler_leakage_pra2013}; such circuit-based leakage reduction algorithms will likely be required in any quantum computing implementation.

Figure \ref{fig:cphase_fidelity}c shows the fidelity of this two-qubit unitary gate $U$ from numerical simulation of the procedure. The fidelity of the unitary gate was defined as 
\begin{equation}
F_U = 1 - \left[ f_1(U_{\mathrm{CZ}}) - f_1(U) \right]^2 - \left[ f_2(U_{\mathrm{CZ}}) - f_2(U) \right]^2
\end{equation}
where $f_1$ and $f_2$ are the two Makhlin invariants \cite{Makhlin_invariants} for two-qubit gates. 
Makhlin invariants are identical for different two-qubit unitary gates if they are equivalent up to single qubit operations.
We find that fidelity better than 99\% is achievable for $\tau_1 \simeq 10$ns, which also leads to very small leakage.  
Figure \ref{fig:cphase_fidelity}d shows the pulse shape of $E_{\mathrm J}^{(1b)}$ and  $E_{\mathrm J}^{(2a)}$ for $\tau_1$=10ns.
The total time duration for the whole process is about 30ns.
In real devices, the fidelity can be lower due to other sources of noise, but here we use only a simple form for the pulse shapes which are not fully optimized as in Refs. \cite{martinis_geller_pra2014,ghosh_geller_pra2013}, so there is some room for improvement.
We also considered Gaussian shape pulses and obtained similar results. 

We can estimate the realistic fidelity of the encoded CPHASE gate constructed here from the fidelities of the $z$-control pulses and the adiabatic process. Since any single qubit logical gate involves at most three rotations (i.e. three pulse steps), the encoded CPHASE gate requires at most 12 pulse steps. Assuming the $z$ pulse fidelity of 0.999 and a fidelity of the adiabatic gate $U$ in Eq.~(\ref{eq:UCPHASE}) betweeen two transmons of about 0.99, the fidelity of the total process can be estimated to be better than $F_2 > 0.999^{12} \times 0.99 \simeq 0.978$. Better optimization or different sequences may improve the fidelity. Of critical comparison, the already demonstrated physical CPHASE gate fidelity of 0.99  \cite{5xmon} also includes a single adiabatic operation and single qubit corrective operations, so the encoded CPHASE gate should be achievable with a similar fidelity. The encoded CNOT gate can be implemented with CPHASE gate and single qubit gates, and we can expect similar fidelity for CNOT gate.

% Fig5
\begin{figure}
  \includegraphics[width=\linewidth]{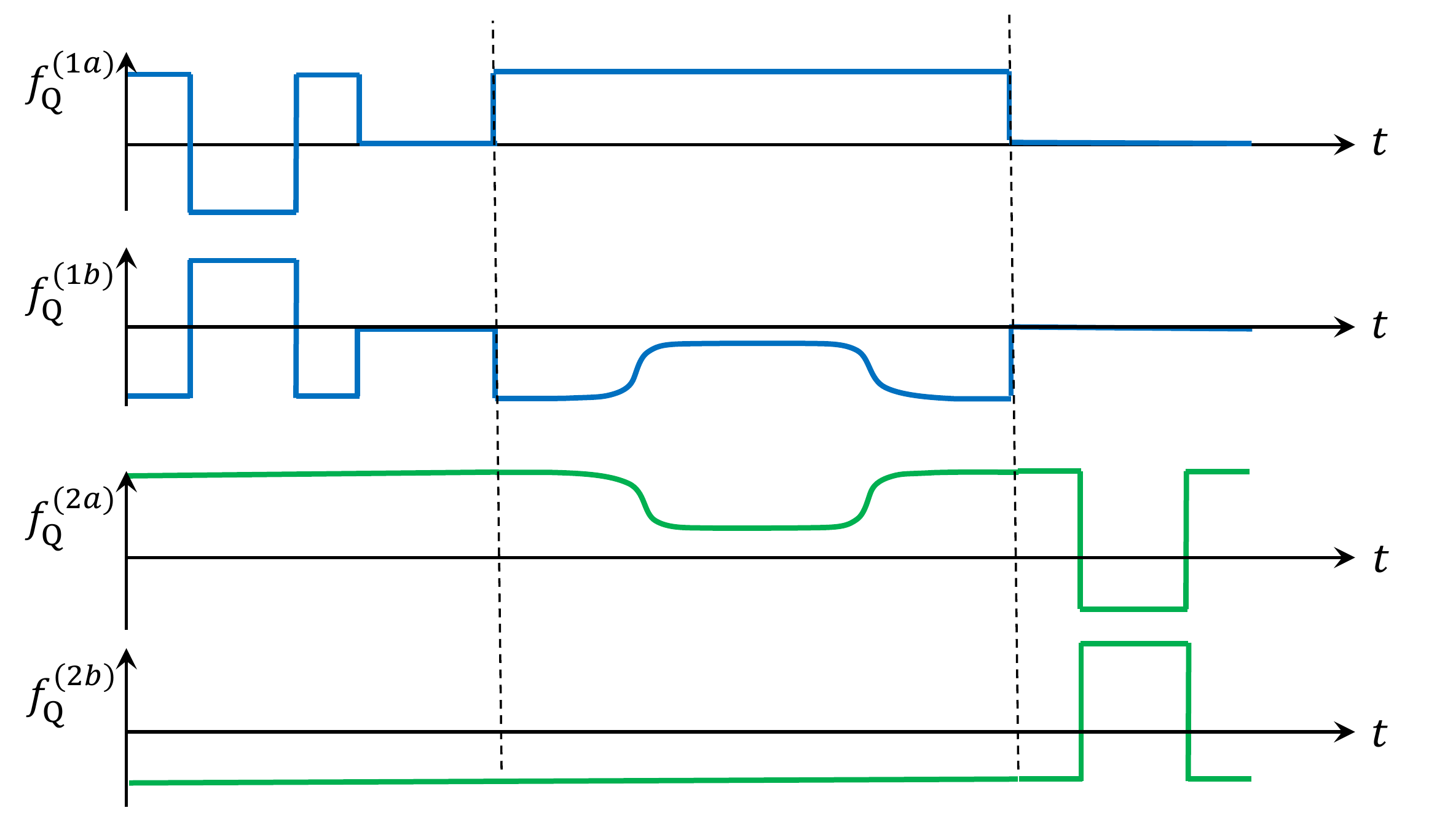}\\
  \caption{{\bf Pulse sequence for encoded CPHASE gate.}  This schematically shows a sequence for CPHASE gate in Eq. (\ref{eq:UCPHASE}). Single qubit phase gate is implemented with three step Euler rotations, and Pauli X and identity gates are implemented as a single rotation. The two vertical dashed lines separate single qubit gates and two-qubit adiabatic operation.}
  \label{fig:cphase_pulse}
\end{figure}

Figure \ref{fig:cphase_pulse} schematically depicts a sequence of DC pulses for the logical CPHASE gate, using the expression in Eq. (\ref{eq:UCPHASE}). The first three pulses in encoded qubit 1 implement a phase gate and the next resonant pulse realizes a Pauli X gate. The second encoded qubit is pulsed to qubit frequencies such that the encoded qubit 2 rotates by $2n\pi$ to implement the identity operation. Then the two-qubit adiabatic gate between transmons $1b$ and $2a$ is applied. After that, an X gate is applied to encoded qubit 1 as a single resonant pulse step and a phase gate is applied to encoded qubit 2 in three rotations. This particular implementation of CPHASE contains only 9 single qubit operations, better than the general 12 single qubit gates we discussed above.   
 
Our choice of encoded qubit is for the sake of simplicity and straightforward incorporation of physical qubit operations into logical gate operations.
We also considered an alternative choice, $\left(|01\rangle \pm |10\rangle \right) / \sqrt{2}$ in the same subspace, which more closely resembles choice for encoded spin qubits.
With this encoded qubit, the constant capacitive coupling leads to a constant energy gap between encoded qubit states, and the $z$ control of each physical qubit allows tunable $\hat{\sigma}^x$ operation. Single qubit logical gates can be implemented in a similar way,
and the adiabatic two qubit operation will need additional single qubit unitary gates to transform to the CPHASE gate due to the basis change of the encoded qubit.  

The capacitive coupling between transmons is typically constant and determined solely by the geometry of the SC islands. 
This coupling is effectively turned on and off by the qubit frequency differences.
With more complicated control circuits as in the gmon architecture \cite{gmon,gmon_theory}, the capacitive coupling can also be tunable and completely turned off, giving a very large on/off ratio. The tunable capacitive coupling removes the need to detune each transmon to avoid unwanted resonances, hence significantly simplifying the qubit frequency controls during the CPHASE operation. This also allows rotating the encoded qubit around any axis in the full $xz$ plane, reducing the necessary rotation steps to two for any single qubit logical gates \cite{two_step_sq_gate}.

\subsection{Initialization}

% Fig6
\begin{figure}
  \includegraphics[width=\linewidth]{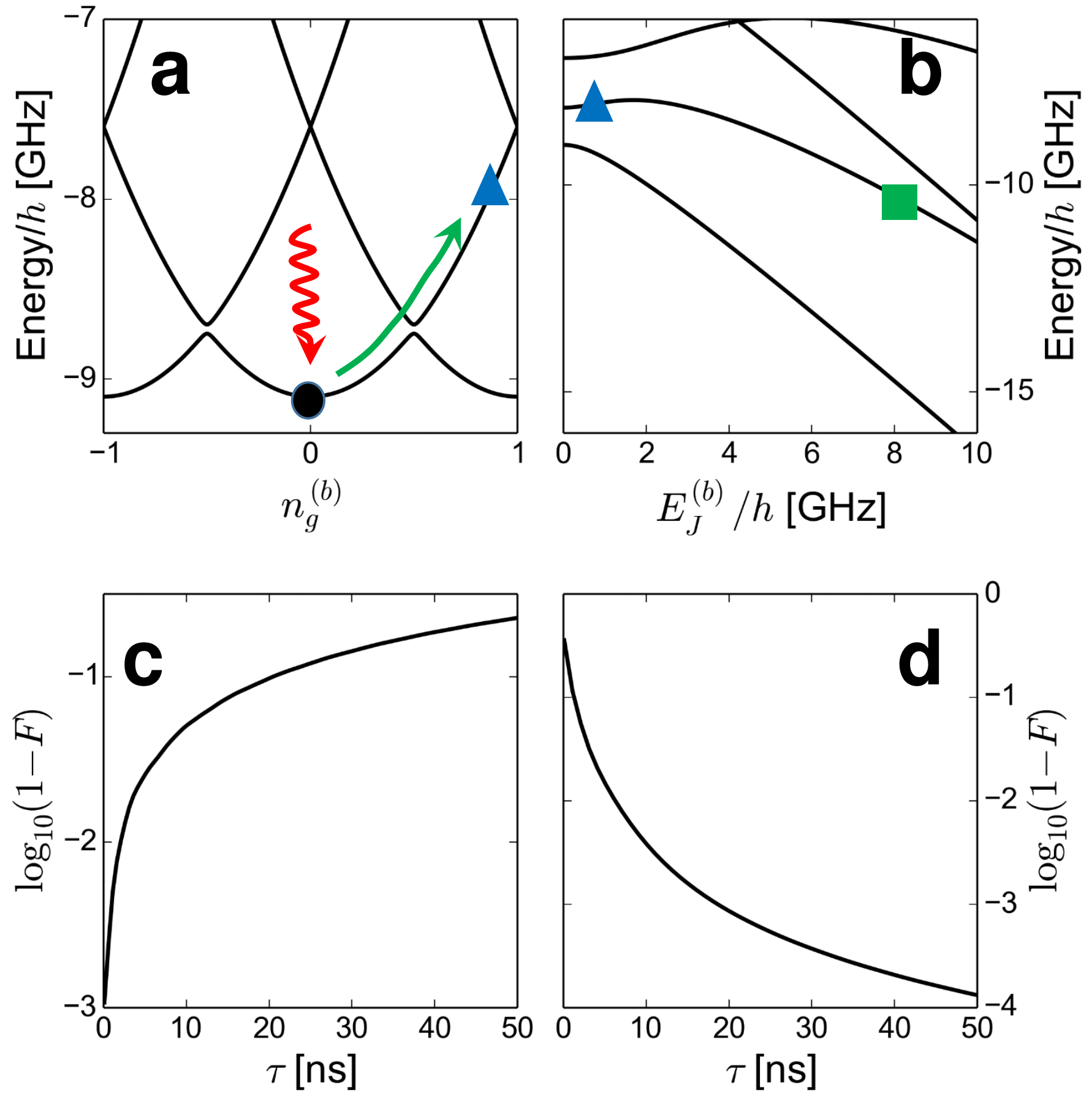}\\
  \caption{{\bf Initialization scheme for the encoded qubit.} {\bf (a)}, the energy spectrum of an encoded qubit with the second transmon in the charge qubit regime. After thermalizing the qubit into the ground state (black dot), $n_{\mathrm g}^{(b)}$ is tuned from 0 to some value between 0.5 and 1. If this change is done fast enough, the qubit is in the first excited state (blue triangle). Then the qubit is adiabatically moved into the transmon regime (green square) by increasing $E^{(b)}_{\mathrm J}$ as shown in {\bf (b)}. {\bf (c)} and {\bf (d)} show the fidelity of these processes as a function of total time duration, respectively.}
  \label{fig:initialization}
\end{figure}

In spin systems the encoded qubit can be initialized fast and with high fidelity by loading pairs of electrons in the singlet state directly from the Fermi sea provided by the leads supplying the quantum dots, then adiabatically separating the singlet into two dots \cite{Petta2005}. Electrons' fermionic and particle nature enables this - a quantum property that may be emulated with engineered many-body photonic systems (for example, Refs. \cite{Greentree_Tahan_nphys2006,Hartman_Brandao_LaserPhontonRev2008}), but which is in no way practical in the near term. One could also engineer a two qubit system where the ground state is a singlet, for example, by making the coupling between the 2-qubits much greater than the qubit splittings (and, e.g,  waiting for relaxation to the ground state). Here, though, one would want to quickly move out of this regime in order to do gates at an implementable speed in addition to turning off as much as possible qubit-qubit couplings, which would be very challenging. Here, we provide an alternative initialization scheme that only requires fast DC pulses. 

The ground state of the two transmon system is $|00\rangle$, which is not in the encoded qubit subspace defined by Eq.~(\ref{eq:logicalQ}). To initialize the system into $|0\rangle_{\mathrm Q}$=$|01\rangle$ without microwave control, we propose using a process analogous to the Landau-Zener (LZ) tunneling \cite{LZ_Landau,LZ_Zener}. For this procedure, we need tunability of the gate charge $n_{\mathrm g}^{(b)}$ of the second transmon, which can be provided by connecting a capacitor with a voltage control to the transmon (see Fig. \ref{fig:system}b) or by using the side-gate for gatemons. The initialization procedure is as follows. First we tune the transmon qubit into the charge qubit regime where $E^{(b)}_{\mathrm J}$ is much smaller than $E_{\mathrm C}^{(b)}$ by tuning $\Phi_\mathrm{ext}$ (or $V_{\mathrm g}$ for a variable super-semi JJ) with $n_{\mathrm g}^{(b)} \simeq 0$. Then, via thermalization (by waiting the relaxation time or by coupling to a dissipative reservoir) the qubit reaches the ground state (black dot in Fig. \ref{fig:initialization}a). (The thermalization could instead be done before tuning to the charge qubit regime.) In this charge qubit regime, the two lowest energy states anti-cross at the sweet spot $n_{\mathrm g}^{(b)}$=0.5. By changing the gate charge $n_{\mathrm g}^{(b)}$ from 0 to a value larger than 0.5, we can induce the LZ tunneling to prepare the charge qubit in the first excited state (blue triangle). Then, we can tune $E^{(b)}_{\mathrm J}$ back to the operating transmon regime ($E^{(b)}_{\mathrm J} \gg E_{\mathrm C}^{(b)}$) [green square in Fig.\ref{fig:initialization}b]. If we tune $E^{(b)}_{\mathrm J}$ exactly to be zero, then there is a crossing instead of anti-crossing, and the fidelity will be much better. But some finite value will be allowable as long as we can change $n_{\mathrm g}^{(b)}$ fast enough. 

Figure \ref{fig:initialization}c shows the calculated fidelity of the LZ tunneling in the charge qubit regime of Fig.\ref{fig:initialization}a as a function of the total time $\tau$ taken to change $n_{\mathrm g}^{(b)}$. Here fidelity is defined as $F$=$| \langle \Psi_{\mathrm{target}}| \Psi_{\mathrm{final}} \rangle |$.
We have used system parameters easily available in real systems, 
$E^{(a)}_{\mathrm C}/h$=$E^{(b)}_{\mathrm C}/h$=375MHz, $E^{(a)}_{\mathrm J}/h$=12GHz, $E^{(b)}_{\mathrm J}/h$=50 MHz, $E_{\mathrm{cc}}$=30MHz.
$n_{\mathrm g}^{(b)}$ was changed from 0 to 0.8. 
As is the case for typical LZ tunnelings, the fidelity is better with faster change of the parameter. We expect to see fidelity better than 99\% for a LZ process of a few ns.
Tuning back to the transmon regime is essentially an adiabatic process, and the fidelity increases with slower change (Fig. \ref{fig:initialization}d). We changed $E^{(b)}_{\mathrm J}/h$ from 50MHz to 8.33GHz, and the fidelity is better than 99\% for a process of a few tens of ns. So this initialization process will take $\sim$ 20 ns to prepare the logical qubit state with fidelity of $\sim$ 99\%. The effect of charge and quasiparticle noise during this process is a concern that should be investigated experimentally, but charge qubits have been shown to have $T_1$ times up to 0.2 ms \cite{Kim_Palmer_prl2011}. Variants of the flux qubit are especially stable to quasiparticle and charge noise fluctuations even at small qubit splittings \cite{fluxmon_Oliver}.

\subsection{Measurement of the qubit states}

Since an encoded qubit is in a state 
\begin{equation}
|\Psi\rangle = \alpha |0\rangle_{\mathrm Q} + \beta |1\rangle_{\mathrm Q} = \alpha |01\rangle + \beta |10\rangle~,
\end{equation}
the encoded qubit can be measured by measuring either of the physical qubits using a standard method, such as dispersive measurement \cite{blais_pra2004,wallraff_nature2004,Schuster_PRL2005,Wallraff_PRL2005} (which can be multiplexed). 
The choice of our encoded qubit in Eq. (\ref{eq:logicalQ}) allows us to translate the single qubit state into the encoded qubit states.
With a choice of a singlet-triplet-like encoded qubit states, $\left(|01\rangle \pm |10\rangle \right) / \sqrt{2}$, the encoded qubit state can also be measured after some single qubit gates are applied to turn them into the encoded qubit states as above, or they could be measured directly by dispersive measurement since these states correspond to different resonator frequencies  \cite{Gambetta_tunable_coupling_transmon_prl2011,Srinivasan_Houck_prl2011}.  

Unlike the spin system where measurement of a singlet can be done electrostatically using a projective measurement \cite{Petta2005}, the dispersive measurement of SC qubits using a transmission line resonator still requires a microwave carrier, which is fine as a proof of concept. We would prefer a measurement approach that takes full advantage of our encoded qubit architecture, with qubit energy completely separated from microwave source. One possibility is to convert the encoded qubit to another quantum system (or measurement qubit) that is long-lived classically but can be read out digitally or with fast base-band pulses (in other words a latched readout), e.g., Ref. \cite{dwave_readout_sst2010}. A compromise option is to do dispersive measurement but still utilize lower bandwidth lines: we can either tune $E_{\mathrm J}$ directly or swap the qubit with another one with a different frequency such that it can be readily measured.

\section{Discussion}

We proposed a scheme for a ``dual rail'' superconducting quantum computer where each qubit consists of two tunable physical qubits. Encoded 2-qubit operations are found to require only a single physical 2-qubit gate and single qubit pulses. Since physical 2-qubit gates are typically much more costly in time and fidelity this means that the overhead of encoded operations as proposed here is not significant, especially compared to spin qubits. 

In this encoded qubit architecture all qubit manipulations are achieved solely by the $z$-control pulse sequences of individual qubits. This removes the requirement of microwave $xy$-control lines necessary in conventional transmon or similar qubit devices, simplifying classical control circuitry significantly. In addition, the encoded approach may allow lower requirements for available bandwidth per line, the potential for less crosstalk, and a reduction in needed timing accuracy as the encoded qubit states are nearly degenerate. Removing the need for microwave control frees the choice of qubit frequency from the cost and availability of microwave electronics. One is then able to design physical qubits with higher (or much lower) frequency that might enable higher temperature qubit operation (which may benefit from work already underway to enable high magnetic field compatible circuits for Majorana experiments \cite{Kouwenhoven2015} in higher-$T_c$ materials) or qubits made from degenerate quantum circuits as in symmetry protected approaches \cite{Doucot_Ioffe_rpp2012,Kitaev_toric_code_AnnPhys2003,Brooks_Kitaev_Preskill_0piQubit_pra2013}, of which there is a natural connection to how spin qubits are inherently protected. 

% leakage reduction
Encoding a qubit in a two-dimensional subspace in a larger Hilbert space introduces leakage error. For our encoded qubits, the relaxation process of individual transmons will lead to leakage out of the encoded qubit space. For a single gate operation such as CNOT of duration $\tau$, the leakage error due to the $T_1$ process would be $1-\exp(-\tau/T_1) \simeq$ 0.04\% for $\tau$=40ns and $T_1$=100$\mu$s, which would slightly reduce the error threshold for quantum error correction \cite{leakage_toric_code}. 
While a single gate operation of a few tens of ns does not lead to significant leakage errors, a long sequence of gate operations in a large system can be a problem. Particularly, a single logical qubit for fault-tolerant quantum computing such as the surface code will consist of many encoded qubits and a logical operation will be a sequence of operations on those encoded qubits. Therefore, leakage reduction units (LRU) \cite{leakage_reduction_unit} will likely be essential. For example, a Full-LRU based on one-bit teleportation \cite{leakage_toric_code} would require an ancilla qubit for each encoded qubit and additional CNOT operations and measurements after each logical CNOT operation. Qubits especially designed for large relaxation times, such as variants of fluxonium \cite{fluxonium_Manucharyan_science2009}, may be particularly promising for our approach (e.g., a $T_1$ time of 1 ms would lead to a leakage error per CNOT of 4$\times 10^{-5}$) and would reduce the overhead for leakage mitigation dramatically.

% fluxmon
The recently demonstrated capacitively-shunted flux qubits \cite{fluxmon_theory,fluxmon_Oliver} (or ``fluxmon") may also provide a promising alternative. They have comparable coherence times and a larger anharmonicity than transmons. Qubit-qubit coupling through mutual inductance would also provide transversal $xx$ coupling like the capacitive coupling between transmon qubits, so the formalism used in this work should be applicable as well. They also offer benefits for initialization as they can be tuned to the flux qubit regime down to very small qubit splitting while being protected to $T_1$ processes that flux qubits offer, and readout can also be done by using a DC SQUID \cite{Bylander_Oliver_nphys2011,Jin_Oliver_PRApplied2015} without a transmission line.
 
% future work
In the next phase of this design philosophy one can consider how to mimic other beneficial properties of spin qubits: very weak coupling between qubit states to charge noise and phonons, a fast and selective two-qubit gate via a Pauli exclusion like mechanism or an interaction that mimics it, very large ON/OFF ratios, and fast initialization via some as yet unknown method.

%% Bibliography

\bibliographystyle{nature}

%\bibliography{references}

%% Acknowledgement
\section{Acknowledgements} 
We thank C. M. Marcus, A. Mizel, W. D. Oliver, B. Palmer, and K. D. Petersson for useful discussions.

% Author contributions
\section{Author contributions}
All the authors contributed to the planning of the project, interpretation of the results, discussions, and writing of the manuscript.
Y.-P.S performed the theoretical and numerical calculations. 

% Additional information
\section{Additional information}
{\bf Competing financial interests:} The authors declare no competing financial interests.

\end{document}